\documentstyle[aps,prd,epsfig]{revtex}
\newcommand{\bee}{\begin{equation}}
\newcommand{\ee}{\end{equation}}
\newcommand{\beea}{\begin{eqnarray}}
\newcommand{\eea}{\end{eqnarray}}

\newcommand{\Tr}{\mbox{Tr}}
\newcommand{\VEV}[1]{\left\langle #1\right\rangle}

\preprint{COLO-HEP-446
}
\begin{document}

\title{A Variant Approach to the Overlap Action}
\author{Thomas DeGrand}
\author{(MILC Collaboration)}
\address{
Department of Physics,
University of Colorado, 
        Boulder, CO 80309 USA
}
\date{\today}
\maketitle
\begin{abstract}
I describe an implementation of the overlap action,
 which is built from an action which is itself an approximate overlap action.
It appears to be about a factor of 15-20 less expensive to use, than
the usual overlap action with the Wilson fermion action as its
kernel. Ingredients include a fat link to suppress coupling to dislocations
and a free field action with a spectrum which resembles an overlap; much
of the gain comes from the use of eigenmodes of the approximate action to
begin the overlap calculation.
 As a physics example,
I compute the quark condensate in finite volume in the quenched approximation.
\end{abstract}
%
%

\section{Introduction}
There has been a great deal of interest recently in lattice fermion actions
which implement an exact chiral symmetry without doubling\cite{ref:reviews}.
The one explicit realization of such an action is the overlap action
of Neuberger\cite{ref:neuberfer}. It obeys the simplest version of the
Ginsparg-Wilson\cite{ref:GW} (G-W) relation.  
All published studies of the
overlap action in four dimensions
\cite{ref:FSU1,ref:HJLu,ref:over4d,ref:FSU,ref:FSUMORE1,ref:FSUMORE2,ref:FSUMORE3,ref:HJL,ref:LIU}
 use the Wilson fermion 
action as their starting point. These actions are apparently very expensive to 
simulate.  However, it is a common expectation that a better starting action
 would simplify the evaluation of an overlap action, if many fewer steps
of iteration would compensate for the increased cost per step of the more
complicated action. That expectation is realized by an action (or family
of actions) I describe here. With it, some physics calculations using an
overlap action in quenched approximations can be done without recourse
 to supercomputers.
 Ingredients in my scheme
 include a fat link to suppress coupling to dislocations
and a free field action with a spectrum which resembles an overlap; much
of the gain comes from the use of eigenmodes of the approximate action to
begin the  calculation of overlap eigenmodes.

These techniques are obviously inspired by the fixed point action program
for constructing classically perfect fermion actions\cite{ref:FP}.
The best action I have found  is, however, not, as far as I know,
 a fixed point action of any renormalization group transformation.

 Versions of these ideas have been presented many times
in two dimensional models\cite{ref:2dim}.

After setting some conventions in Sec. \ref{sec:notat},
I will describe the candidate actions and their tests in
 Sec. \ref{sec:actions}. I will then present in Sec. \ref{sec:condense}
 a calculation of the
 quark condensate in finite volume with one of the new actions, along the
lines of the calculation of Ref. \cite{ref:HJL}.

\section{Notation and Conventions}
\label{sec:notat}
I will call a generic lattice Dirac operator $d$ and the
 overlap Dirac operator $D$. The eigenvalues of the simplest implementation
of a G-W action lie on a circle of radius $r_0$ and the
(massless) overlap Dirac operator is
\bee D(0) = r_0(1+ {z \over{\sqrt{z^\dagger z}}} )
\label{eq:gw}
\ee
where $z = d(-r_0)/r_0 =(d-r_0)/r_0$ and $d(m)=d+m$ is the massive
 Dirac operator for mass $m$
 (i.e. $r_0$ is equivalent to a negative mass term.)
The overall multiplicative factor of $r_0$ is a useful convention; when
 the Dirac operator $d$ is thought of as ``small'' and Eqn. \ref{eq:gw} is
expanded for small $d$, $D \simeq d$.
Apart from this overall factor of $r_0$, 
my conventions are those of Ref. \cite{ref:FSU}.

The Hermitian Dirac operator for mass $m$ is defined as $h(m)= \gamma_5d(m)$
and the overlap Hermitian Dirac operator is denoted as $H(m)$. Specifically,
\bee
H(0) = r_0(\gamma_5 + \epsilon(h(-r_0)))
\label{eq:hgw}
\ee
where $\epsilon(x)$ is the step function, $\epsilon= -1$ if $x<0$,
$\epsilon=1$ if $x>0$. I will refer to the argument of the step function
as the ``kernel'' of the overlap.

It is convenient to define the massive overlap Dirac operator as
\bee
D(m) = (1- {m\over{2r_0}})D(0) + m
\ee
and then the squared massive Hermitian Dirac operator is
\bee
H(m)^2 = m^2 + (1- {m^2\over{4r_0^2}})H(0)^2.
\ee
The zero eigenvalue eigenmodes of $H(0)$ are chiral, with 
$\langle \phi | \gamma_5 |\phi \rangle = \pm 1$, and the nonzero
eigenvalue eigenmodes of $H(0)$ come in pairs of equal and opposite
eigenvalues. For these modes,
$\langle \phi_\lambda | \gamma_5 |\phi_\lambda \rangle = 
\langle \phi_\lambda | H(0) |\phi_\lambda \rangle /(2r_0) = \lambda/(2r_0)$.

In a background gauge field carrying a topological charge $Q$, $D(0)$ (and
$H(0)$) will have $Q$ pairs of real eigenmodes with eigenvalues 0 and $2r_0$.
In computing propagators (for example for $\VEV{ \bar \psi \psi}$),
it is convenient to clip out the eigenmode with
 real eigenvalue  $2r_0$, and to define the
subtracted propagator as
\bee
\tilde D(m)^{-1} =
 {1 \over {1 - {m\over{2r_0}}}}[ D(m)^{-1} - {1\over {2r_0}}].
\label{eq:dm}
\ee
\section{The overlap with a better kernel}
\label{sec:actions}
\subsection{Ingredients}
A good $d$ or $h$ for use as a kernel in Eq. \ref{eq:gw} or \ref{eq:hgw} should
already ``look like'' $D$ or $H$.  This means that the eigenvalues of its
eigenmodes should lie (approximately) on a circle and its low lying eigenmodes
should be approximately chiral--in a spectroscopy calculation, the additive
renormalization of the bare quark mass, as measured, for example, through
the variation of the pion mass with bare quark mass, should be small.
Most implementations of the overlap action in the literature use the
Wilson action, which does not satisfy either of these criteria:
the eigenvalues of the free Wilson action sit like beads on a string around
a set of four circular arcs (with real parts of their eigenvalues ranging
 from 0 to 8), and the additive mass renormalization of the interacting
theory is $\Delta am_q \simeq 1$ for simulations at lattice spacings near
0.15 fm.
Use of the clover action instead of the Wilson action improves the
chiral properties but does nothing for its (free field) eigenmode spectrum.

Improving the action involves two ingredients.

First, to improve the chiral properties of the action, thin links are
replaced by fat links. Actions with fat links are
already quite chiral as shown by their small mass renormalization and
 $Z_A \simeq 1$ \cite{ref:FAT,ref:FATg}. In this work I have  studied
fat link actions with two blockings:  The first uses
 APE-blocking\cite{ref:APEblock}:
\beea
V^{(n)}_\mu(x) = &
(1-\alpha)V^{(n-1)}_\mu(x) \nonumber  \\
& +   \alpha/6 \sum_{\nu \ne \mu}
(V^{(n-1)}_\nu(x)V^{(n-1)}_\mu(x+\hat \nu)V^{(n-1)}_\nu(x+\hat \mu)^\dagger
\nonumber  \\
& +  V^{(n-1)}_\nu(x- \hat \nu)^\dagger
 V^{(n-1)}_\mu(x- \hat \nu)V^{(n-1)}_\nu(x - \hat \nu +\hat \mu) ),
\label{APE}
\eea
with  $V^{(n)}_\mu(x)$  projected back onto $SU(3)$ after each step, and
 $V^{(0)}_\mu(n)=U_\mu(n)$ the original link variable.
I have mostly studied a large amount of fattening, $\alpha=0.45$, $N=10$,
but have looked at the smaller value $\alpha=0.25$, $N=7$.
The second blocking is a ``hypercubic'' blocking devised by
A. Hasenfratz\cite{ref:AHblock}.
The fat links of this action are confined to a hypercube, so this
action is more local than the APE-blocked actions.
The mean plaquette at $\beta=5.9$
 for (0.45,10) APE blocking is about 2.98,
for (0.25,7) APE blocking it is 2.88,
and for the hypercubic blocking,  2.84.

I believe that what is important here is that the fat gauge links decouple the
fermions from short distance structure in the gauge field, not that the links
are fattened in a particular way.

\begin{figure}[thb]
\epsfxsize=0.8 \hsize
\epsffile{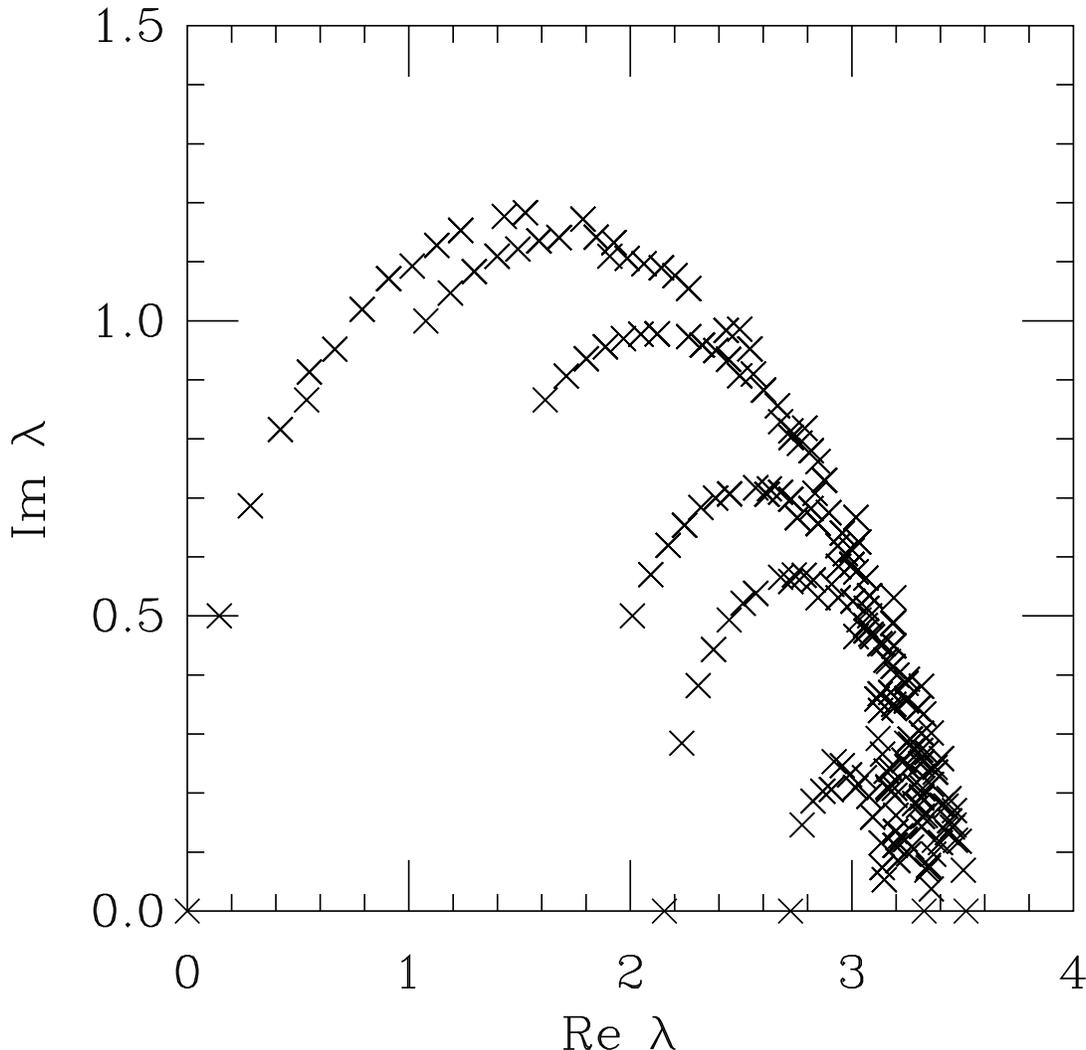}
\caption{
Eigenmode spectrum for the free field planar action on a finite lattice.
 Only the positive
imaginary part modes are shown.
}
\label{fig:pl0}
\end{figure}

The eigenvalues of a GW action  lie on a circle. I determine
the best $d$ by taking a free field test action and
 varying its parameterization to optimize its eigenvalue spectrum
 (in the least-squares
sense) for circularity, for some $r_0$. It was convenient to let the
parameter $r_0$ also be a free parameter. The simplest extension of a nearest
neighbor action which can have its eigenvalues lying approximately on a circle
is ''planar:'' it has
scalar and vector couplings
 $S=\sum_{x,r}\bar\psi(x)(\lambda(r) +i \gamma_\mu \rho_\mu(r))\psi(x+r)$
for $r$ connecting nearest neighbors ($\vec r=\pm\hat\mu$;
$\lambda=\lambda_1=-0.170$, $\rho_\mu^{(1)}=-0.177$) and diagonal
neighbors ($\vec r=\pm\hat\mu \pm\hat\nu$, $\nu\ne\mu$;
$\lambda=\lambda_2= -0.061$, $\rho_\mu^{(2)}= \rho_\nu^{(2)}= -0.0538$.
 The constraint
 $\lambda(r=0)= -8\lambda_1 -24 \lambda_2$
 enforces masslessness on the spectrum,
and $-1 = 2 \rho_\mu^{(1)} + 12\rho_\mu^{(2)}$ normalizes
the action to $-\bar \psi i \gamma_\mu \partial_\mu \psi$ in
the naive continuum limit.  The corresponding value of $r_0$ is
1.6.
 A plot of the free field eigenvalues
on a finite lattice is shown in Fig. \ref{fig:pl0}.
Presumably the places where its eigenvalues are purely
 real (but not zero) correspond in the continuum limit to theories with
various numbers of massless free fermions, exactly as for the Wilson action.

The action also includes a clover term, with its coefficient set to
the tree-level value appropriate to this action of $C_{SW}=1.029941$.
Thus $d$ has no $O(a)$ artifacts. All the links, including the ones in the
clover term, are replaced by fat links. The gauge connections to the
diagonal neighbors are the average of the two shortest path connections.
The cost of the action is about 6.5 times that of the usual clover action.

While it is probably not germane to the present discussion, the dispersion
relation of this action is improved compared to that of clover or
 Wilson actions.

No claims are made for uniqueness--or even optimality.
The optimizations do not have to be done with enormous precision since the
overlap formula itself deforms almost any kernel action into a chiral action.
The amount of fattening is also a free parameter, and there will almost
certainly be tradeoffs between ease of implementing the overlap and the
desire that the action show good scaling behavior. At large levels of
fattening, there appears to be little renormalization of the parameters
of a free action, and so little tuning is required to produce actions
which behave well in simulations. Thus, the overlap parameter $r_0$ will
be kept at its free field value of 1.6.

\subsection{Testing actions}
In what follows I will focus on comparisons of the overlap action with the
Wilson action as a kernel, the ``Wilson overlap'' and with the fat link planar 
action as a kernel, or ``planar overlap.''
I have looked at six actions in all. Most tests involve the
Wilson overlap or the (0.45,10) APE-blocked planar overlap.
I have briefly investigated an overlap action with the
(0.45,10) APE-blocked fat link clover action as a kernel.
I have also investigated planar overlaps with (0.45,10) APE-blocked links
and with hypercube-blocked links. Finally, I tested the ``Gaussian''
action, which has been proposed as a candidate
approximate FP action\cite{ref:FATg}.

  The step function is approximated
by either the polar formula introduced by Neuberger
\bee
\epsilon(z) \rightarrow \epsilon_N(z) = 
z\frac{1}{N}
\sum_{j=1}^N \frac{1}{c_j z^2 + s_j}
\label{eq:polar}
\ee
($c_j=\cos^2(\pi(j+1/2)/(2N))$, $s_j=\sin^2(\pi(j+1/2)/(2N))$),
or by a fourteenth order Remes polynomial, following the work of
Edwards, Heller and Narayanan\cite{ref:FSU}.  In practice, the polar
formula works very robustly for the planar overlap, when I rescale
$h(-r_0)$ to $h(-r_0)/r_0$ in Eq. \ref{eq:polar} and take $N=6$ to 10.
The Wilson overlap is much more delicate and I have used the
 Remes approximation exclusively for it. I rescaled the operator $h(-r_0)$
by a factor of 1/2.5 to map its eigenvalues into a range where the
 Remes algorithm has small errors.  The inverses of the terms in the sum of
Eq. \ref{eq:polar} are found using a multi-mass conjugate gradient 
routine\cite{ref:multimass}.

\begin{figure}[thb]
\epsfxsize=0.8 \hsize
\epsffile{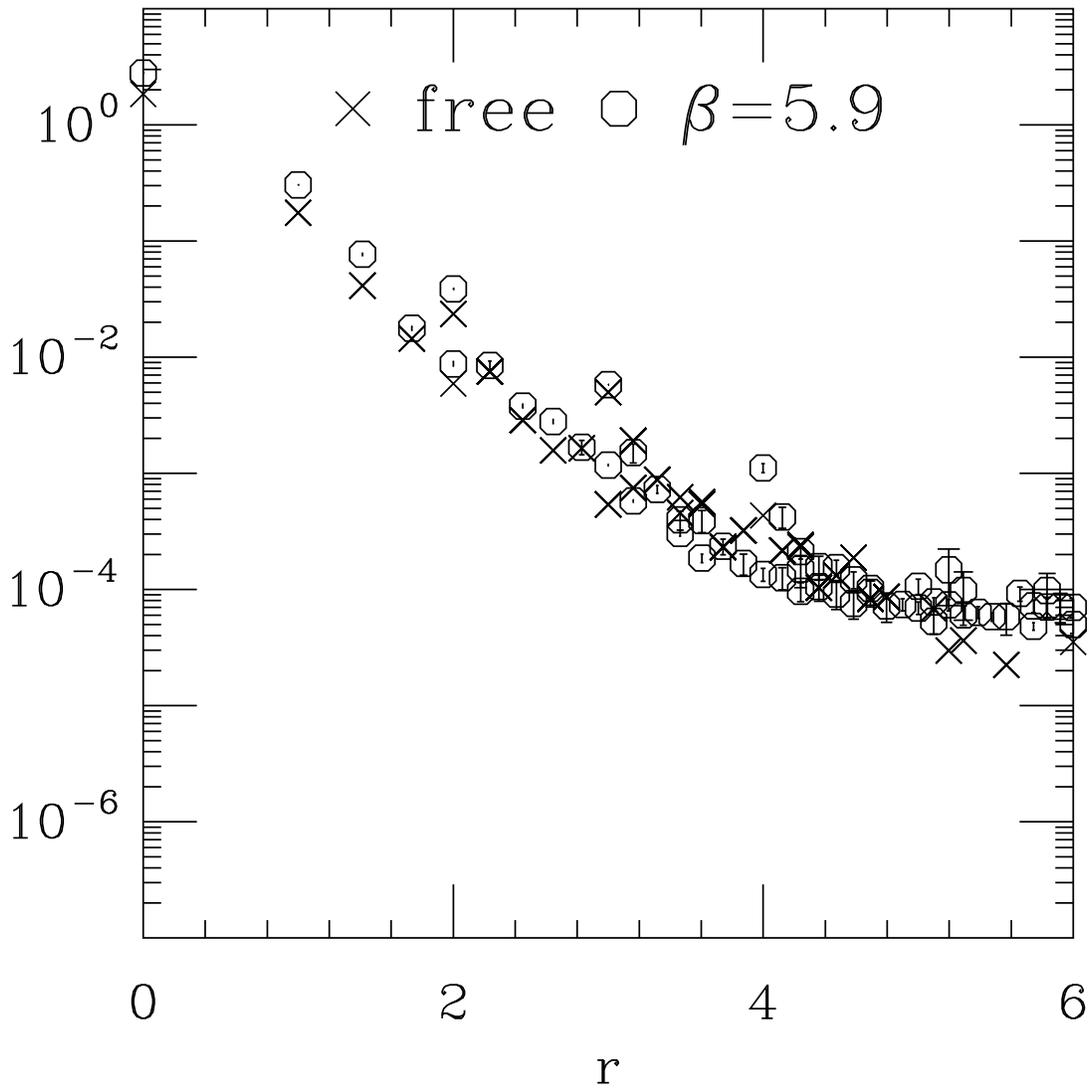}
\caption{
Range of the Wilson overlap action, in free field theory and on a set of
$\beta=5.9$ $8^4$ lattices.
}
\label{fig:ftwo}
\end{figure}

\begin{figure}[thb]
\epsfxsize=0.8 \hsize
\epsffile{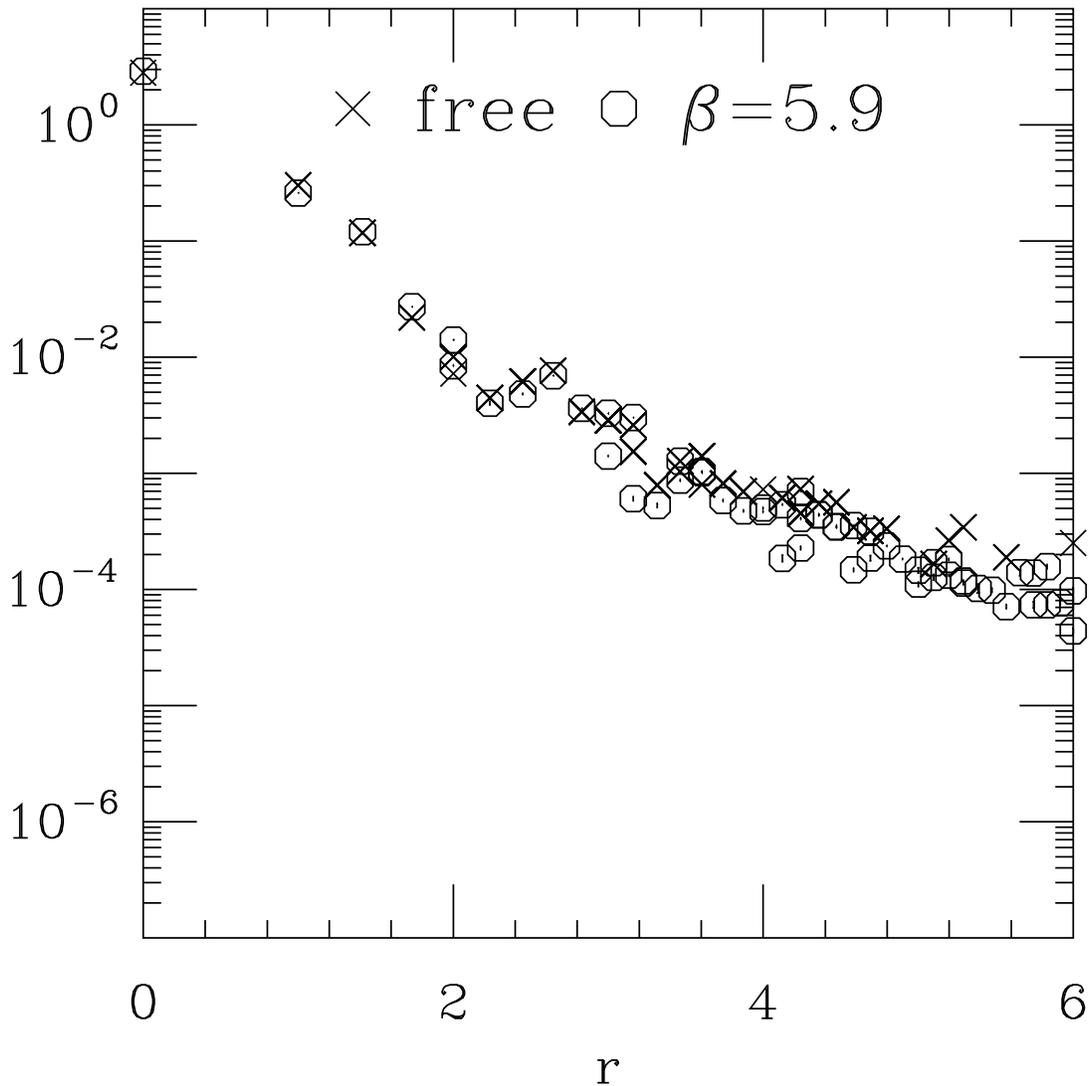}
\caption{
Range of the (0.45,10) APE-blocked
 planar overlap action, in free field theory and on a set of
$\beta=5.9$ $8^4$ lattices.
}
\label{fig:fthree}
\end{figure}

Let's begin by looking at the actions. The range of the action is computed
using a variation of the calculation of  H\'ernandez, Jansen, and
 L\"uscher\cite{ref:HJL}:
 Shown in Figs. \ref{fig:ftwo} and \ref{fig:fthree} is a comparison of
Wilson and planar overlap  $\sqrt{ |D(0)\chi|^2 }$
for a delta-function source $\chi$ at the origin, as a function of
distance $r = \sqrt{x_\mu^2}$, for the free field action
 as well as on a set of
$\beta=5.9$ $8^4$ configurations.  The two actions show similar exponential
falloff with distance. Of course, these pictures do not show anything
about the locality of the gauge connections in the actions, just the
fermions. In perturbation theory\cite{ref:BD} the fat link action has a cloud 
of glue of size $\sqrt{\alpha N/3}$ convoluted over every fermion offset.

It might be relevant that the plaquette for the fat link action has
a value $\Tr (1-U_p)/3) \simeq 1 - 2.97/3 = 0.01$, which should be sufficient
for the argument of H\'ernandez, Jansen, and L\"uscher\cite{ref:HJLu}
to insure the locality of the overlap action.

All my tests of the overlap begin with finding the eigenmodes of $H$
with the smallest eigenmodes (actually  the  eigenmodes of $H$ diagonalized
in a basis which is composed of the smallest eigenmodes of $H^2$).
Low lying eigenmodes of $h(m)$ and $H(0)$ are found using an adaption of
a conjugate gradient algorithm of Bunk et. al.
and  Kalkreuter and Simma\cite{ref:eigen};
I modified a code originally written for staggered fermions by K.~Orginos.

I have implemented most of the standard tricks for efficient evaluation of
eigenmodes of
$H$. I compute a set of the $N_0$ low lying eigenmodes of $h(-r_0)$
(typically $N_0=10-20$), and project them out during the calculation of
the operator $\epsilon(h(-r_0))) \chi$.

\begin{figure}[thb]
\epsfxsize=0.8 \hsize
\epsffile{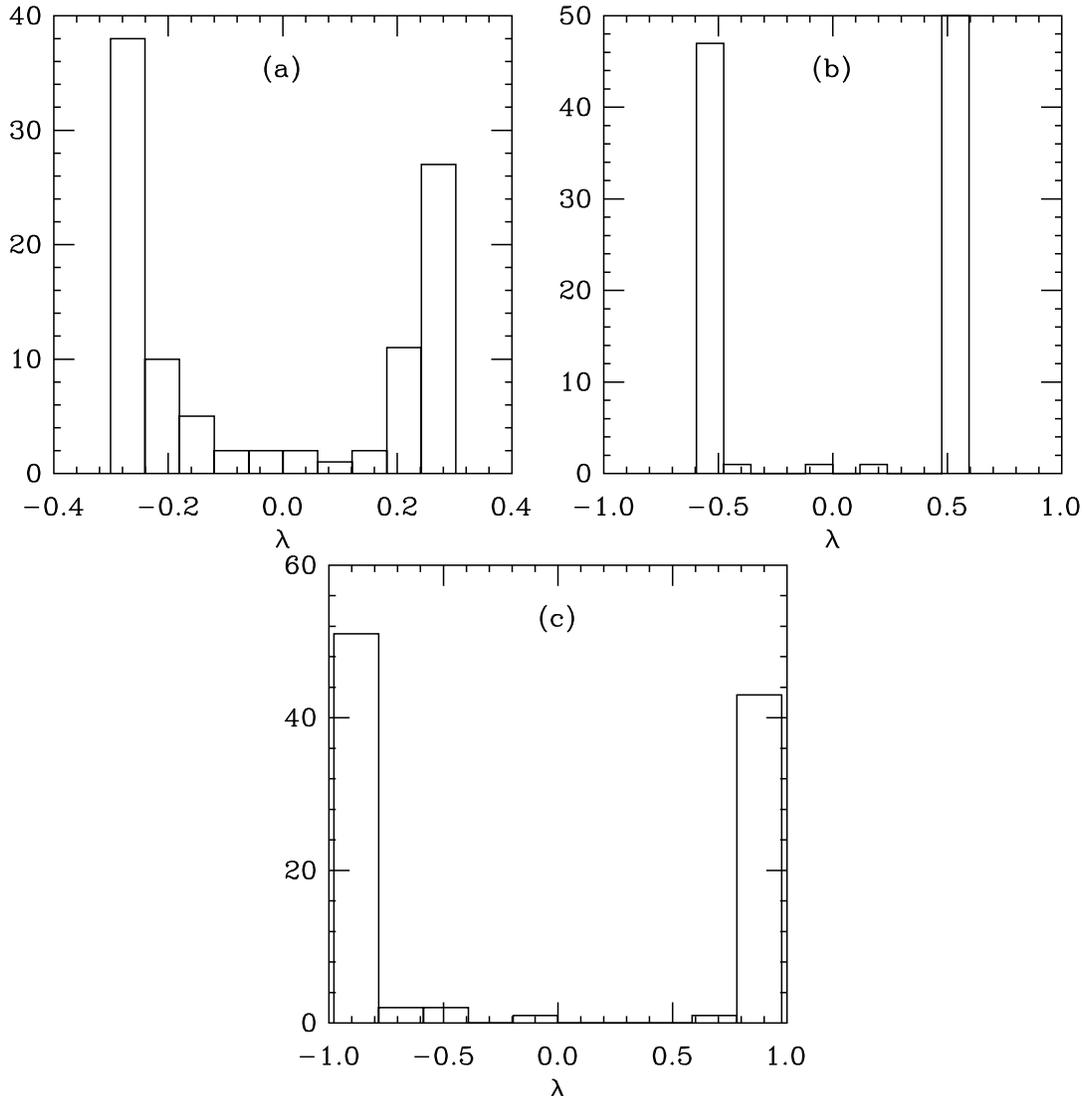}
\caption{
Spectrum of $h(-r_0)$ for the ten smallest eigenvalues of $h(-r_0)^2$
for (a) the Wilson action ($r_0=1.65$), (b) the fat link planar action
$(r_0=1.6)$, and the fat link clover action ($r_0=1.0$), on 10 $8^4$
$\beta=5.9$ configurations.
}
\label{fig:rhos}
\end{figure}

A first sign that the fat link will make the overlap better behaved comes
from looking at the eigenmodes of the kernel function $h(-r_0)$. Fig.
\ref{fig:rhos} shows a histogram of the ten smallest eigenmodes of $h(-r_0)$
reconstructed from the ten smallest eigenmodes of $h(-r_0)^2$ on a set
of ten $8^4$ $\beta=5.9$ configurations.  The fat link actions have
many fewer small eigenmodes. Eigenmodes near zero are associated with
 dislocations, gauge configurations on which topological objects are about
to disappear. It is difficult for approximations to the step function to 
process these modes. The absence of these modes in the spectrum of $h(-r_0)$
is why the polar form of $\epsilon_N$ can perform well, even for small $N$.
There is a lot of discussion in the literature (see for example Ref.
\cite{ref:vranas}) about improving overlap or domain wall fermions by using
gauge actions which have fewer dislocations than the Wilson gauge action,
but a fermion action which could not see dislocation would work just as well.
A fat link action does just that. While the fat link
clover action is as much as an
 improvement as the planar action, it has a much higher conditioning number
than the planar action. This is reflected in the amount of work needed to
find the eigenvalues (and will also affect the number of inner
 conjugate gradient steps needed to evaluate the step function): the
average number of Rayleigh iterations needed to collect the ten lowest
eigenmodes is 3600 for the Wilson action, 5100 for the fat link clover action,
and 1900 for the (fat link) planar action.

Because $[H(0)^2,\gamma_5]=0$, eigenmodes
 of $H(0)^2$  can be simultaneously eigenmodes of $\gamma_5$. All eigenmodes
are constructed in a basis of chiral eigenfunctions. In that basis,
eigenmodes of $H(0)^2$ are expected to be doubly degenerate. This
degeneracy can be used to test algorithms.
A low order  polar approximation to the step function
 does a poor job of producing
degenerate pairs of eigenmodes of $H(0)^2$ for the Wilson overlap
and at this level of test it must
be discarded.

The calculation of the condensate in Sec.~\ref{sec:condense} requires
the quark propagator $D(m)^{-1}$, with the contribution from chiral
 modes removed. I do this by taking the set of eigenmodes of $h$
and finding the two lowest eigenmodes $H(0)^2$ in a basis containing
one state of each chirality. The chirality of the highest state identifies
the chiral sector which has no zero modes. I then perform the inversion
of $H^2=D^\dagger D$ in this sector. This calculation is also
accelerated by finding and projecting out some
 large number ($O(10)$) of eigenmodes of $H(0)^2$ during this inversion.

A trick which I have not noticed to have been  emphasized in the literature
is actually the source of much of the gain in efficiency of the planar overlap.
This is to begin the calculation of eigenmodes of $H$ by first finding
a set of low lying eigenmodes of $h(m)$ for some potentially useful $m$,
and beginning the calculation for $H$ using the eigenmodes of $h$ (rather
than, say, beginning with a set of random vectors).
The combination of the small number of small eigenmodes of $h(-r_0)$
for the fat link action and the use of good trial functions makes the
planar overlap quite efficient compared to the Wilson overlap.

Most of my tests are on a set of $8^4$ $\beta=5.9$ lattices. Let us consider
a set of examples which illustrate the differences. In the planar
overlap I take the $N=10$ polar approximation to the step function,
scale the argument by $r_0=1.6$, and project out $N_0=10$ eigenmodes of
$h(-r_0)$ in the step function.  The Wilson overlap uses the fourteenth-order
Remes approximation, scales its argument by 1/2.5, and projects 20 eigenvalues
of $h(-r_0)$.
In both cases I wish to find the lowest two eigenmodes of $H(0)^2$,
regardless of chirality, and I begin by finding and utilizing the four
smallest eigenmodes of $h(0)$. I investigate a configuration which happens
to have topological charge $Q=1$.
 The planar overlap calculation needs 109
Rayleigh iterations and 2529 inner conjugate gradient steps to find the two
lowest modes
(at a cost of 6.5 times the equivalent Wilson action step). The Wilson overlap
needs 1220 Rayleigh iterations, and 86,723 inner conjugate gradient steps,
to reach the same level of accuracy (about $10^{-5}$ for the eigenvalues).
Had we wanted the ten smallest eigenmodes of $H(0)^2$, using 20 trial
 eigenmodes of $h(0)^2$, the cost would be 306 Rayleigh iterations and
 8189 inner conjugate gradient
 steps
for the planar overlap, or about 26 inner steps per Rayleigh iteration,
while the Wilson overlap uses 5820 Rayleigh iterations and a million
inner conjugate gradient steps (about 170 inner steps per iteration).
This is a savings in computer time of about a factor of eighteen for the
planar overlap compared to the Wilson overlap. On other configurations,
the number of inner CG's per Rayleigh step
for the planar overlap ranges between 20 and 30, and
the number of Rayleigh iterations ranges from 300 to 500.

\begin{figure}[thb]
\epsfxsize=0.8 \hsize
\epsffile{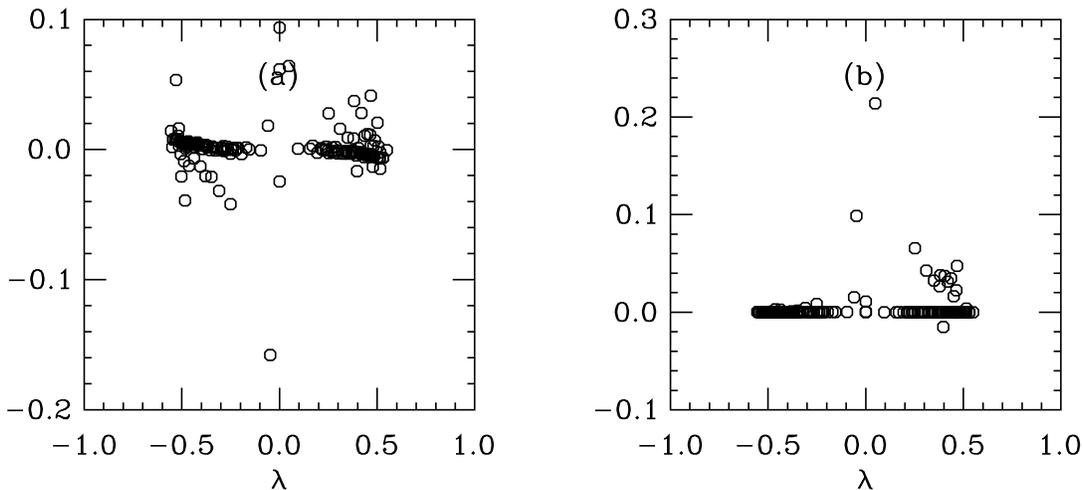}
\caption{
Differences between exact eigenvalues of $H(0)$ and (a) eigenvalues of the
planar $h(0)$ and (b) eigenvalues of $H(0)$ produced by diagonalizing
$H(0)$ in the lowest 20 eigenvalue eigenmodes of $h(0)^2$, as a function
of overlap eigenvalue.
}
\label{fig:diffs4510}
\end{figure}

\begin{figure}[thb]
\epsfxsize=0.8 \hsize
\epsffile{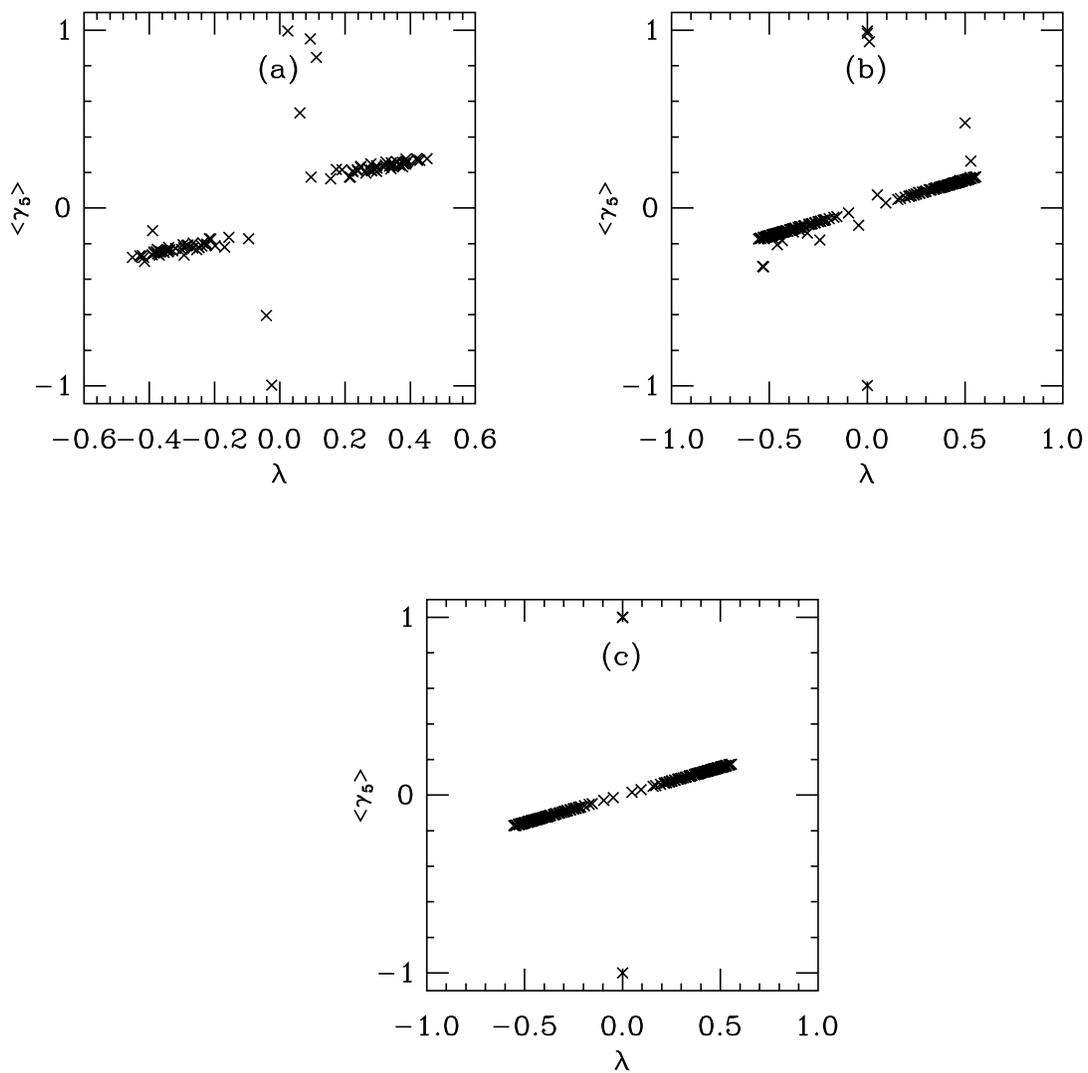}
\caption{
$\VEV{\gamma_5}$ vs. eigenvalue for the planar overlap: (a) 20 smallest
eigenvalues of $h(0)$, (b) 20 smallest eigenvalues of $H(0)$ from a
diagonalization of the basis of (a), (c) 20 smallest eigenvalues
of $H(0)$ from a full calculation, all on a set of ten $\beta=5.9$ $8^4$
lattices.
}
\label{fig:gg}
\end{figure}

\begin{figure}[thb]
\epsfxsize=0.8 \hsize
\epsffile{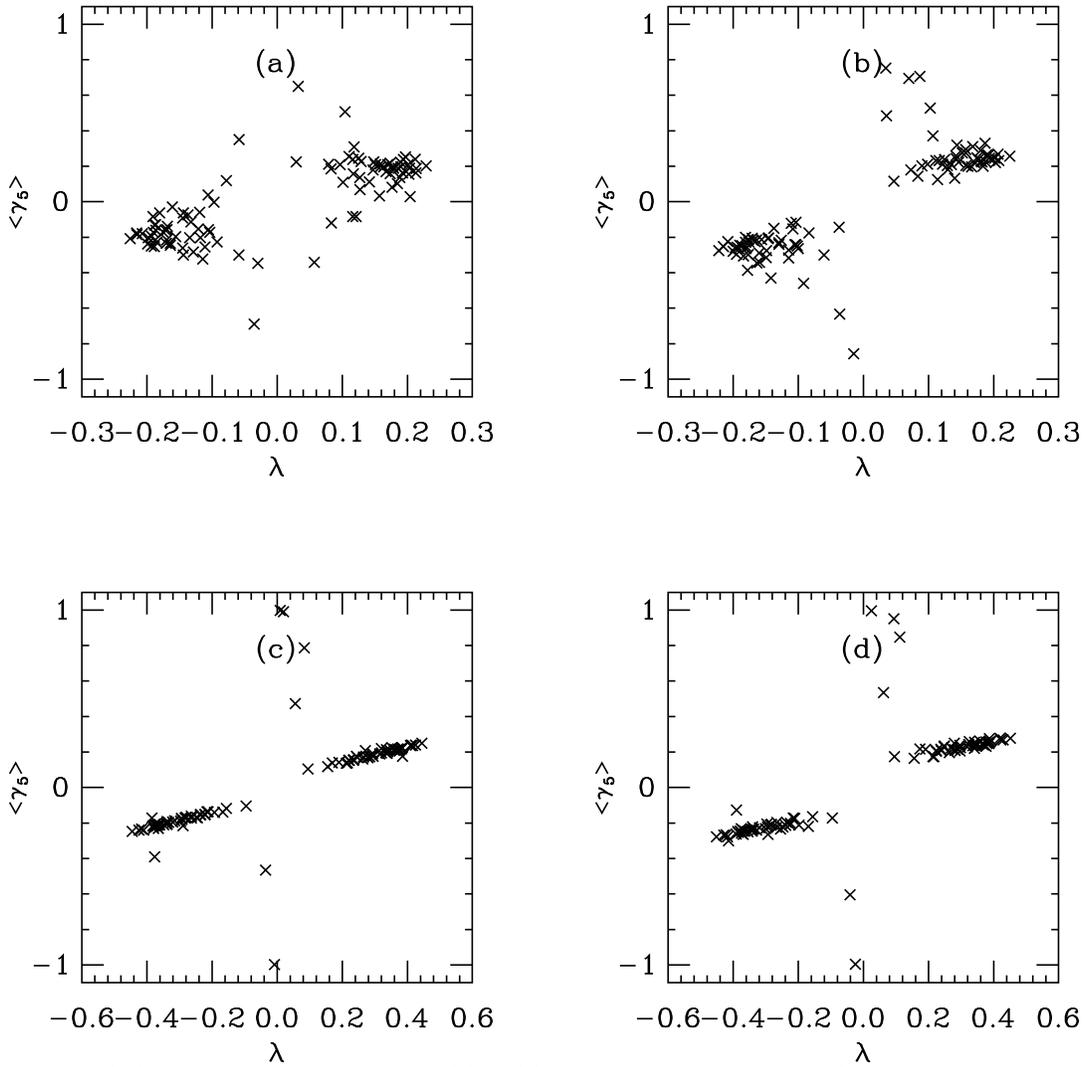}
\caption{
$\VEV{\gamma_5}$ vs. eigenvalue for several $h(0)$'s.
(a) thin link Wilson action, (b) non-perturbative thin link clover action,
(c) Gaussian action, (d) planar action.
}
\label{fig:g5}
\end{figure}

The difference in the number of inner conjugate gradient steps
is due to the fact that the Wilson $h(-r_0)$ has many more small eigenvalues
than the fat link planar
 action. The difference in the number of Rayleigh iterations
arises because the initial trial vectors, eigenmodes of $h(0)$, are
close to being eigenmodes of $H(0)$. I illustrate this with a scatter
plot of the change in an eigenmode, plotted as a function of the value
of the overlap eigenmode, in Fig. \ref{fig:diffs4510}.
 A graph of $\VEV {\gamma_5}$ vs.
eigenmode for this action, at various stages of the overlap
calculation, is shown in Fig. \ref{fig:gg}. It is also useful to look at the
scatter of chirality vs. eigenvalue for several trial actions which
might be candidates for an overlap kernel. This is shown in Fig. \ref{fig:g5}.
I show $\VEV {\gamma_5}$ vs.
eigenmode for the Wilson action, non-perturbative thin link clover
action (with $C_{SW}=1.85$), fat link planar action, and fat link
Gaussian action.
 From this picture, one would suspect that the thin
link actions would not provide good trial wave functions for the overlap.

\begin{figure}[thb]
\epsfxsize=0.8 \hsize
\epsffile{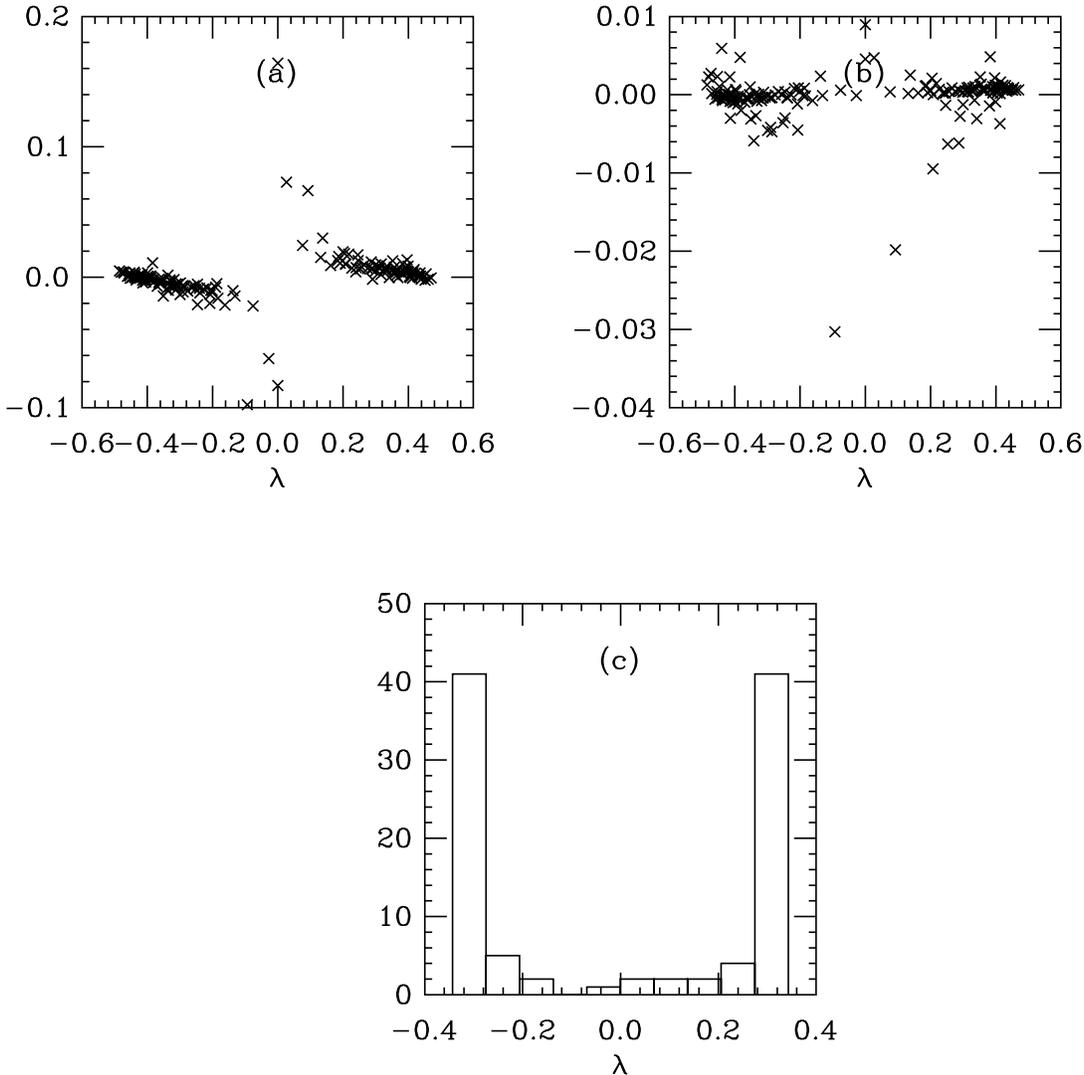}
\caption{
$\VEV{\gamma_5}$ vs. eigenvalue for the planar action and its overlap with
(7,0.25) fattening: (a) 20 smallest eigenvalues of $h(0)$,
(b) 20 smallest eigenvalues of $H(0)$ from a diagonalization of the basis of
(a), (c) histogram of ten smallest eigenmodes of $h(-r_0)$,
on $\beta=5.9$ $8^4$ configurations.
}
\label{fig:a257}
\end{figure}

I have tested another fat link action, a planar action with 7 APE-blocking
steps and $\alpha=0.25$. I kept $r_0=1.6$ since tests of the
 fat link clover action with this level of fattening at $\beta=5.9$
also showed little additive mass renormalization.
The number of small eigenvalues of $h(-r_0)$ increases slightly compared to the
(0.45,10) case, and the number of inner CG steps grows, correspondingly,
to about  about 30.  But the eigenmodes of $h(0)$ do not seem to be
 as good a trial
basis as they were for the fatter link action, and the number of
Rayleigh iterations grows from about 10,000
 to 12,000-15,000 for the same calculation as was
done above. A set of pictures showing relevant information for this
action is shown in Fig. \ref{fig:a257}. This action does not see some of the
 instantons
that the (0.45,10) fattening saw. That is reflected in the scatter
of the eigenvalues. 

One might hope that a better trial function would help. A simple way
of changing the trial wave function is to vary the mass $m$ in $h(m)$.
Trials at $m= -0.05$ and $-0.1$ did not produce any dramatic changes.
This is an obvious place for future work.

The planar action with hypercubic fat links shows nearly identical
behavior. Both these actions need about fifty per cent more inner CG
steps than
the (0.45,10) planar action but are still a large improvement over
the Wilson overlap--and one could still run them on work stations.
They might be useful in contexts where a very fat link might have
undesirable properties.

Finally, the Gaussian action with (0.45,10) APE blocking requires about
half as many inner CG calls as the (0.45,10) planar action, typically
3000-5000. However, this action requires about 2.5 times as much storage
and 2.5 times as much CPU time per inner CG
step as the planar overlap action, since
its couplings span a hypercube (80 neighbors rather than 24). Thus it
does not seem to be competitive with the planar action without clever
coding \cite{ref:BERNstuff}.

The (0.45,10) fat clover action, with $C_{SW}=1$, was also briefly investigated
as a kernel for the overlap. It seemed to need 100,000-300,000 inner CG
steps in the fiducial calculation, in 2500 to 8000 Rayleigh iterations,
beginning from $m=0$ trial wave functions. (There is an extreme
variation in these numbers from lattice to lattice tested.)
 Eigenmodes of the fat link clover
action are apparently quite different from those of the corresponding
overlap action.
The number of inner CG steps reflects the greater
  conditioning number of $h(-r_0)$.

These tests are certainly incomplete. It may be that I have simply written
a very inefficient Wilson overlap, though most of the code for the two
methods is common. My Wilson overlap is so expensive that it cannot be tuned,
whereas it is easy to test variations of the planar overlap.
  The ``design philosophy'' of beginning with an
action which is close to an overlap action certainly produces actions
which are inexpensive enough to run on small computers
 (work stations in my case).

\section{The quark condensate in finite volume}
\label{sec:condense}
In order to do a little physics in what is otherwise a paper about technique,
I present a calculation of the chiral condensate in
the quenched approximation
 with the (0.45,10) fattened planar overlap
 action.  The method is (almost) exactly that of the pioneering calculation
of Hernandez and Jansen and Lellouch\cite{ref:HJL}: one computes the
condensate in background gauge field configurations of fixed topology
labeled by winding number $\nu$\cite{ref:verbar}. The condensate
is
\bee
\Sigma_\nu = m \Sigma^2 V(I_\nu(m\Sigma V)K_\nu(m\Sigma V) + 
I_{\nu+1}(m\Sigma V)K_{\nu-1}(m\Sigma V)) +  {\nu \over {m V}}
\label{eq:sigma}
\ee
where $m$ is the quark mass, $V$ is the volume, and $\Sigma$ is the 
infinite volume condensate. $I_\nu(z)$ and $K_\nu(z)$ are modified
Bessel functions. In practice, one computes $\Sigma$ ``without topology,''
by working in the chiral sector which has no zero eigenmodes (and doubling
the result).
 Then one measures
$\Sigma_\nu - \nu/(mV)$. In practice, the lattice number needs an
additive renormalization: in zero mass, this is removed by the replacement
of $D^{-1}$ by $\tilde D^{-1}$ as in Eq. \ref{eq:dm}. There can also
be contribution which vary with the quark mass, so the lattice number
will need to be fit to
\bee
\VEV{\bar \psi \psi}_{sub} = {1\over V} \sum_x \tilde D_{x,x} =
(\Sigma_\nu - \nu/(mV)) + C m +\dots
\ee
One small difference between my calculation and that of Ref. \cite{ref:HJL}
is in the normalization of the propagator: they do not have the prefactor
$1/ (1 - {m\over{2r_0}} )$ of Eq. \ref{eq:dm}.
It is hard to believe that the condensate should not be an odd function of
the symmetry breaking term (the quark mass), and the data does not show
anything but a simple linear dependence on the mass.

Finally, $\Sigma$ is scheme-dependent, and an overall lattice-to-continuum
regulator needs to be computed.

Warned by the results of 
Ref.\cite{ref:HJL} , I restricted my calculation to the $\nu=\pm1$
sector. I generated a set of 40 lattices at each of three volumes $8^4$,
$10^4$, and $12^4$ at $\beta=5.9$. I filtered them to find candidate
$\nu=\pm1$ configurations using a pure gauge measurement of topological
charge\cite{ref:annatop}, which I had previously calibrated against
a set of 10 $8^4$ lattices. I checked that these lattices in fact had
$\nu=\pm 1$ during the calculation of the condensate; only one lattice
failed this test. This obviously leaves out configurations which the fermion
observable identifies as carrying topological charge, but the gauge observable
does not, but the alternative is to process every configuration through
the overlap program. I was left with a set of 10 $8^4$, 13
$10^4$, and 9 $12^4$ lattices on which I computed $\VEV{\bar \psi \psi}$.
I computed propagators at five quark masses, $am= 0.001$, 0.002, 0.005, 0.01,
and 0.02 (again using a multi-mass Conjugate Gradient solver).
I used twelve random sources per lattice (though I blocked data together
before averaging).
This needed about 40, 70, and 125 steps at the three volumes, at
about 20-30 inner CG's per outer step, to reach a fractional squared
residue of $10^{-12}$. As a check, I also calculated $\Sigma$ from the
GMOR relation
$\sum_x \VEV{\pi(x) \pi(0)}= \Sigma/m$;
 as expected in the overlap it reproduced the
direct calculation of $\Sigma/m$ to within $10^{-7}$ (see Ref. \cite{ref:FSU}).

\begin{figure}[thb]
\epsfxsize=0.8 \hsize
\epsffile{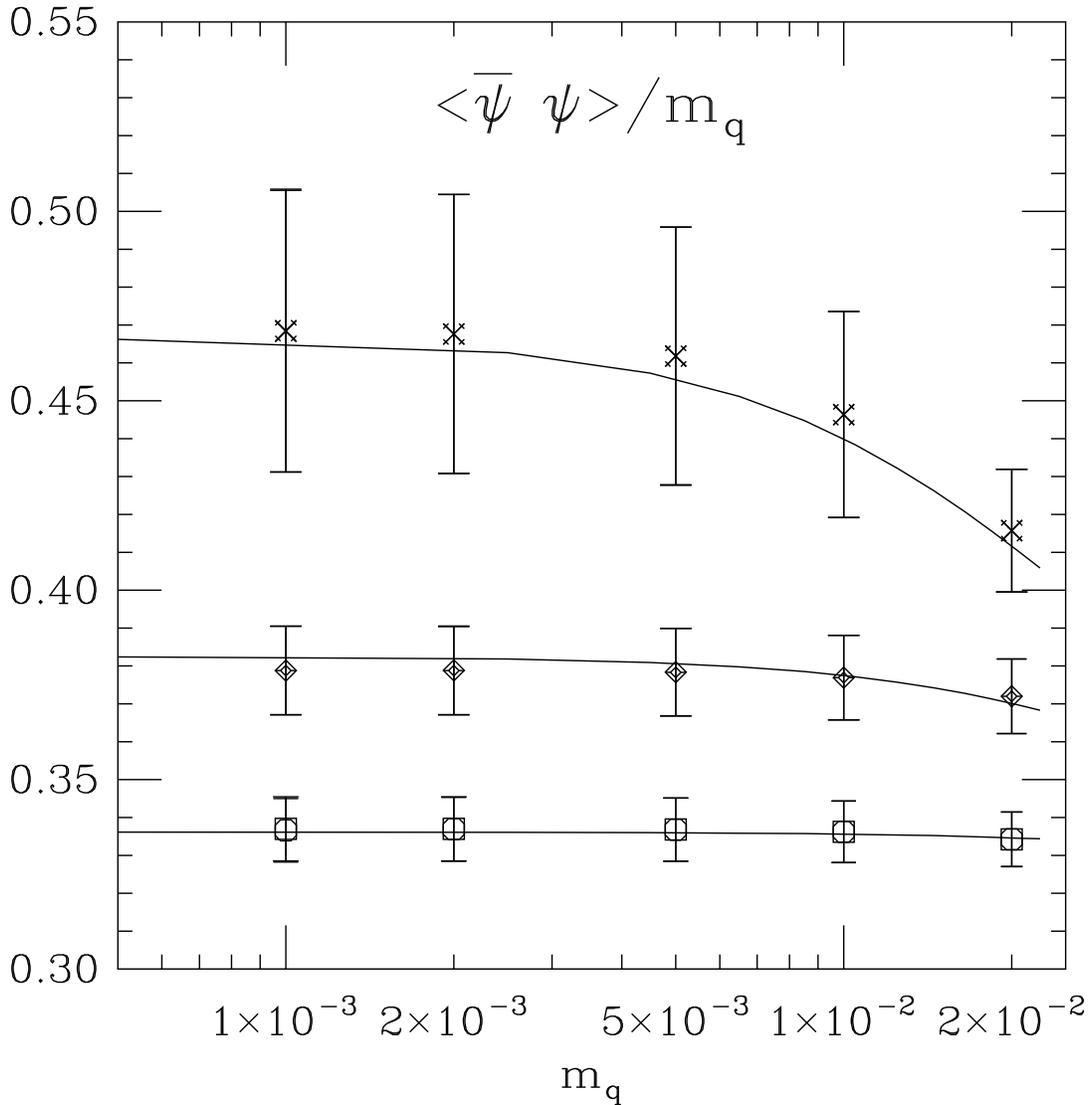}
\caption{
The quark condensate in finite volumes (divided by the quark mass).
From the bottom, the volumes are $8^4$, $10^4$, and $12^4$, The curves
are fits to Eq. \protect{\ref{eq:fit}}.
}
\label{fig:pbp}
\end{figure}

My lattice results are shown in Fig. \ref{fig:pbp}.

The data at each volume are quite correlated. A single-elimination jackknife
fit to
\bee
\VEV{\bar \psi \psi}_{sub} = 
m \Sigma^2 V(I_1(m\Sigma V)K_1(m\Sigma V) + 
I_2(m\Sigma V)K_0(m\Sigma V)) + C m 
\label{eq:fit}
\ee
gives (re-inserting the lattice spacing)
 $\Sigma a^3 = 0.00394(16)$, $C=0.304(8)$. The lattice number is about two
standard deviations higher than the number reported by Ref. \cite{ref:HJL}
at $\beta=5.85$ of $\Sigma a^3 =.0032(4)$. We can sharpen this disagreement
by trading the lattice spacing $a$ for the Sommer radius $r_0$, using
the interpolating formula of Ref. \cite{ref:precis}:
 $\Sigma r_0^3 = 0.215(27)$ for Ref. \cite{ref:HJL}, 
 $\Sigma r_0^3 = 0.354(14)$ here.

A real comparison requires computing the $Z$ factor. I have not done that
yet, but I can make a heuristic attempt at the calculation, by exploiting
the fact that perturbation theory for fat link actions becomes simple in the
limit of large fattening.  This argument is implicit in the discussion
of fat link perturbation theory in Ref. \cite{ref:BD}.
Since it falls outside the main thrust of the paper, I relegate the
discussion to an Appendix.

Taking $r_0$ to be 0.5 fm,
$\Sigma_{\overline{MS}}(\mu = 2$ GeV) = 0.0244(10)(37) GeV${}^3$.
The two errors are from the lattice fit and from an assumed lattice spacing
uncertainty of five per cent\cite{ref:FNAL}.

This calculation produces a number which disagrees badly with
a calculation of the condensate using clover fermions and the GMOR
relation, $\Sigma_{\overline{MS}}(\mu = 2$ GeV) = 0.0147(8)(16)(12) GeV${}^3$
\cite{ref:guisti}.
It is done at smaller lattice spacing and has completely different systematic
uncertainties. Oddly enough, however, my result
agrees well with an earlier analysis of data by Gupta and Bhattacharya:
\cite{ref:GB}
$\Sigma_{\overline{MS}}(\mu = 2$ GeV) = 0.024(2)(2) GeV${}^3$
(statistical and lattice spacing uncertainties).

Of course, all the potential weaknesses of this calculation, reported by
Ref. \cite{ref:HJL} apply here, too: the lattices are small,
and the lattice spacing is large. And it is a quenched calculation.

All of the computations of $\Sigma$
 were done over about a month of running on a set
of four or five  450 Mhz Pentium II work stations.
\section{Conclusions}
None of the results shown here are particularly surprising, but that does
not mean that they might be totally devoid of interest.
A good approximate overlap action is easier to convert into an exact overlap,
than any random action like the Wilson action. Crucial ingredients seem
to be some kind of gauge connection which suppresses dislocations, a free
action which ``resembles'' a free field overlap action, and the use of
as much information from the trial $h(m)$ as possible to begin the calculation
of the overlap action. One should be able to do better.
\section*{Acknowledgements}
I am deeply indebted to Urs Heller for many conversations about the overlap
action, to Robert Edwards for a table of Remes coefficients,  to Kostas
Orginos for a copy of his eigenvalue code,  to Anna Hasenfratz for
discussion about blocking, and to Archie Paulson for help coding. This work was supported by the
U.~S. Department of Energy.

\appendix
\section{Z-factors for very fat actions}

The evaluation of the Z-factor is  straightforward. I will use many of
the techniques of Ref. \cite{ref:LM}, but collect them here for completeness.
We expect $\Sigma(\mu)_{\overline{MS}} = Z(a\mu)\Sigma(a)_{latt}$
with $Z_S = 1/Z_m$ and $Z_m$ is the quark mass renormalization factor. In  
 perturbation theory, $Z_m = 1 + b z$, and
\bee
z =\Delta_{latt} - \Delta_{cont} = 6 \ln(1/(\mu a)) + x,
\ee
where $b = \alpha_s(q*)/(3\pi)$. I choose to use
 $\alpha_s =\alpha_{\overline{MS}}(q^*)$. I will use the plaquette to define
a coupling $\alpha_V(3.41/a)$ (=0.16054 at $\beta=5.9$),
 which will then be converted using two-loop perturbation theory
to $\alpha_{\overline{MS}}(q^*)$, to give 
$\Sigma(\mu= 1/a)_{\overline{MS}}$. I will then run this result to $\mu=2$ GeV,
using the (inverse of the) two-loop running formula for the $\overline{MS}$
quark mass.

$z$ is the difference between a lattice integral
\bee
\Delta_{latt} = \int_k I_{latt}F_N;
\ee
and a continuum integral
\bee
\Delta_{cont} = \int_{k,c} I_{cont}
\ee
I have factored the lattice integrand into a thin link piece
$I_{latt}$
  and  the form factor of the fat quark-gluon vertices $F_N$; for APE-smeared
links  $F_N = (1 - \alpha \hat q^2 /6)^{2N}$
 where $\hat q^2 = \sum_\mu (4/a^2) \sin^2(k_\mu a/2)$.
I am using the notation $\int_k = \int d^4 k / (2\pi)^4$  over the hypercube,
$-\pi/a < k_\mu < \pi/a$; $\int_{k,c}=\int d^4 k / (2\pi)^4$.
Introducing a gluon mass $\lambda$ to regularize its IR divergence,
 $\Delta_{cont} = 5/2 - 3 \ln(\lambda^2\mu^2)$ (in $\overline{MS}$).
 Adding and
subtracting continuum-like terms to isolate the divergence
 in the lattice integral, we write
\beea
\Delta_{latt} = & \int_k( I_{latt} - 3 {\theta(\pi^2 - k^2) \over k^4})F_N
 - 3 \int_k {\theta(\pi^2 - k^2) \over k^4}( 1 - F_N) + 3 \int_k 
{\theta(\pi^2 - k^2) \over k^4}
\nonumber  \\
= & J -3D + 3 \ln{\pi^2 \over { a^2 \lambda^2}}.
\label{eq:magic}
\eea
The first two terms ($J$ and $D$) are IR and UV finite.
Now the point is that $D$ is a continuum-like integral, and all the
dependence of $\Delta_{latt}$ on  the lattice action is in the first term.
But if the fattening is large, the form factor $F_N$, is nonzero only
at tiny $k$, where it is unity. At that point, for all practical purposes,
$I_{latt} \simeq 3/k^4$, and so we expect that $J$
will be small. All the  dependence of $Z_m$ on 
the lattice action, other than its form factor,
 vanishes when we  set $J=0$ in Eq. \ref{eq:magic}.
  $q^*$ (defined according to the prescription of
Lepage and Mackenzie\cite{ref:LM}) is calculated by a similar procedure.

While I can't check this approximation for the overlap action, 
at (0.45,10) fattening, a complete calculation
for clover fermions
gives $x= -5.58$ and $q^*=1.07$, while the $J=0$ result is
$x= -5.44$ and $q^*=1.10$. A similar argument would predict that finite
renormalization factors, such as the vector and axial current renormalizations,
are unity, which is a good approximation to what is seen at large fattening.
 Of course, for small fattening, this
approximation fails badly and a better (non-perturbative?) calculation is
necessary. (For example, the (0.25,7) APE-blocked clover action has
$x=-0.975$, $q^*=2.88$; the approximate result is $x= -2.49$, $q^*=1.59$.)

\end{document}